\documentclass[acmtocl,acmnow]{acmtrans2m} 
\usepackage{graphicx} 
\usepackage{amsmath} 
\usepackage{amsfonts} 
  \newtheorem{theorem}{Theorem}
  
\usepackage{bbm}
\usepackage{enumitem}

\newcommand{\real}{\mathbbm{R}}
\usepackage{mathrsfs,mathtools}
\usepackage{mathtools} 
\newcommand{\dist}{\xrightarrow[H_0]{{\cal D}}}
\newcommand{\njt}{\ensuremath{n_{1 \cdot}}}
\newcommand{\ndt}{\ensuremath{n_{2 \cdot} }}
\newcommand{\ntj}{\ensuremath{n_{\cdot 1} }}
\newcommand{\ntd}{\ensuremath{n_{\cdot 2} }}
\newcommand{\njj}{\ensuremath{n_{11} }}
\newcommand{\njd}{\ensuremath{n_{12} }}
\newcommand{\ndj}{\ensuremath{n_{21} }}
\newcommand{\ndd}{\ensuremath{n_{22} }}

\title{Bayesian Estimation and Regularization Techniques in Categorical Data Analysis}

\author{JAN KALINA\\The Czech Academy of Sciences, Institute of Computer Science, Prague\\\& Charles University, Faculty of Mathematics and Physics, Prague}

\begin{abstract}
This paper explores Bayesian estimation for categorical data, focusing on simple yet effective models that provide a foundation for applying more advanced methods accurately and reliably in real-world applications. We begin by revisiting Bayesian estimators for the binomial distribution and investigating their properties. Next, we develop hypothesis tests for categorical data (sign test, homogeneity test, symmetry test) based on regularized maximum likelihood estimates of the probabilities. Finally, we formulate regularized versions of common association measures for contingency tables and study the regularized version of mutual information, particular for the situation where the regularized version can effectively handle zero counts.
\end{abstract}

\category{62H17, 62F15}

\terms{Categorical data, Bayesian estimation, Hypothesis tests, Regularization}

\begin{document}

\bibliographystyle{acmtrans}

\begin{bottomstuff}
This work is supported by the project 24-11146S (``Maximal entropy portfolio'') of the Czech Science Foundation.
\end{bottomstuff}

\maketitle

\section{Introduction} 

Bayesian thinking, with its origins in \cite{bay}, has recently penetrated many applications in various fields.
Bayesian estimation is an important approach in the analysis of categorical data \cite{joh}, particularly for contingency tables (two-way frequency tables) that aggregate counts according to different categorical variables \cite{agr}.
Analyzing complex Bayesian models requires understanding the basic principles of simpler models \cite{lof}, as well as the principles for selecting appropriate prior distributions \cite{tuy}.

This motivates us to revisit the binomial distribution and two-way contingency tables as simple models for categorical data, and to consider Bayesian estimation of their parameters along with regularized versions of hypothesis tests or association measures.
The broader aim is to develop tailor-made tools for reliable analysis of complex categorical data, increasingly relevant in modern applications \cite{lin}.
Such methods are also relevant for machine learning, where regularized Bayesian estimation can improve the accuracy and robustness of predictions, similarly to how neural networks learn complex patterns \cite{want}.

Common types of regularized estimators are known to be connected to Bayesian thinking~\cite{jay}. 
The recent boom of regularized estimators has focused mainly on continuous data, often 
aimed at sparse methods for variable selection in classification~\cite{led} or covariance matrix estimation~\cite{kaladac}. 

For categorical data, the motivation for regularization may be quite different. 
Regularization can improve robustness to mismatches, i.e., situations when a measurement 
is erroneously assigned to the wrong category (e.g., confusing success and failure for binary data). 
It also reduces the variability of estimates, because maximum likelihood estimates 
have too high a variability for small samples; introducing a small bias can remarkably reduce this variability~\cite{soh}. 
Moreover, regularization directly helps mitigate numerical instability in test statistics, parameter estimates, association measures, and mutual information, improving the reliability and robustness of inference. 
Regularization is particularly useful when some counts in the contingency table are very small 
(or even 0), especially if the number of categories of one or more variables is large. 
While this paper does not address such high-dimensional settings, they represent an important direction for future research.

Regularized estimators often take the form of shrinkage estimators. 
For Bayesian tools, it is natural to obtain the shrinkage estimators as shrunken versions of the maximum likelihood estimator (MLE) towards the mean of the prior distribution.
If the categorical data are in several different groups, then the shrinkage is natural to be considered towards the MLE evaluated across the groups.
Recent proposals of regularized methods for categorical data include hypothesis tests~\cite{wanr}, clustering \cite{bae}, logistic regression models \cite{min}, or monitoring categorical processes \cite{wank}. 
The connection of regularization and Bayesian estimation was investigated for continuous data in \cite{zho}. 
Regularized tools (mainly test statistics) for analyzing categorical data were used for the labor market segmentation in~\cite{tol}. 
Much more intensive attention has been paid to regularized test statistics for continuous data, e.g. for the Hotelling's $T^2$~test~\cite{iss}.

In current machine learning, pattern discovery or association rule learning represents an important task \cite{gol}. So far, individual applications exploiting regularized approaches considered mainly continuous data.
For example, the learned patterns were used to construct classification rules in \cite{bra} or to search for latent clusters in \cite{li}. Rare examples of pattern discovery on categorical data are cited in the overview \cite{sub}
for the integration of genomics or metabolomics data.
It is therefore natural to assume that effective pattern discovery in contingency tables should rely on suitable regularized estimation tools, as these are able to stabilize inference in the presence of sparse counts and improve the reliability of detected associations.

The concept of entropy plays a central role in modern quantitative finance, where it underpins portfolio construction methods designed to remain stable under model uncertainty. In this context, maximal entropy approaches aim to distribute weights as evenly as possible while respecting structural constraints, thus promoting diversification and robustness. A similar rationale applies in the analysis of categorical data: when faced with sparse or zero counts, entropy-based reasoning naturally leads to regularized estimators that stabilize inference by shrinking toward more balanced distributions. Entropy provides a conceptual link between applications in finance and the analysis of categorical data. 
In both cases, entropy-based reasoning encourages balanced and robust estimates, 
which is particularly valuable when counts are sparse or zero~\cite{gup}.

This paper presents a methodological application of regularized Bayesian estimation for categorical data, focusing on handling challenges such as small or zero counts.
Section~\ref{sec:ber} is devoted to the estimation of the parameter of the binomial distribution. There, regularized estimators obtained as Bayesian tools with different choices of the prior distribution are studied and discussed.
Section~\ref{sec:hyp} studies commonly used hypothesis tests for categorical data for the situations with regularized maximum likelihood estimates of probabilities for individual cells.
We formulate regularized versions of some common association measures for contingency tables in Section~\ref{sec:mea} and study the regularized version of mutual information in Section~\ref{sec:ham}.
Section~\ref{sec:con} concludes the paper.

\section{Bernoulli distribution} 
\label{sec:ber}

Let us consider i.i.d. random variables $X_1, \dots, X_n$ coming from the Bernoulli distribution (alternative distribution, 0-1 distribution) with an unknown parameter $\pi \in [0,1]$.
In this section, we recall several well-known versions of the Bayesian estimator of~$\pi$.
These estimators have the form of regularized (shrinkage, penalized) versions of the maximum likelihood estimator.
We provide a novel and rigorous formal treatment of these estimators in Section~\ref{sec:prop}.

It is common for the binomial distribution to perceive $\pi$ as the probability of success and $1-\pi$ as the probability of failure. The sum $\sum_{i=1}^n X_i$ follows the binomial distribution ${\sf Bi}(n,p)$.
The most prominent Bayesian estimators of~$\pi$ will be discussed and compared here.
The MLE obtained as $\hat{\pi} = \frac{1}{n} \sum_{i=1}^n X_i$ is a consistent and unbiased estimator of $\pi$ \cite{rao}. 
From the theory of maximum likelihood estimation, it follows that $\hat{\pi}$ achieves the Rao-Cram\'er lower bound, however only under the assumption of $n \to \infty$.
In a variety of applications, the MLE is not a~holy grail and it may be convenient to deviate from the MLE especially for data with a~small $n$.

Bayesian estimators of $\pi$ typically have the form of regularized estimators as convex linear combinations combining the MLE with some prior information according to
\begin{equation}
\tilde{\pi} = \lambda \frac{\sum_{i=1}^n X_i}{n} + (1-\lambda) {\sf E}\pi, \quad \lambda \in [0,1],
\label{e:gen}
\end{equation}
where ${\sf E}\pi$ is the expectation of the prior distribution of $\pi$.

When cell counts are small, simple continuity corrections such as adding 0.5 are sometimes used to stabilize estimates. 
In this paper, we adopt a Bayesian approach that generalizes this primitive method: by introducing a prior distribution, the posterior mean effectively shrinks the cell probabilities in a principled way, providing improved variance reduction and numerical stability while retaining interpretability. 
This offers a flexible framework that extends the idea of ad hoc continuity corrections to a fully probabilistic setting.

Throughout the paper, the regularization parameter $\lambda$ is considered fixed.
This is in contrast to certain Bayesian choices, where the corresponding $\lambda$ may implicitly depend on $n$ (typically with $\lambda_n \to 1$ as $n \to \infty$). 
Our asymptotic results should therefore be interpreted under the assumption of a fixed $\lambda$.
Throughout the paper, we consider the Bayesian estimates to have the form of the mean of the posterior distribution (if not stated otherwise).

Replacing $\hat{\pi}$ by a~biased estimate is especially appealing if $\hat{\pi}$ is quite unexpectedly near~0 or~1, even if there is no evidence (prior belief, expectation) that it should be so.
Also, the set of all possible values of the MLE, i.e.~$\{0, 1/n, \dots, (n-1)/n, 1\}$, does not allow for a more delicate estimation of $\pi$ for a small $n$, which may justify to consider small deviations from $\hat{\pi}$.
Thus, we can say that the shrinkage towards~${\sf E}\pi$ makes a~correction for the finite sample estimation. This is especially true when $\hat{\pi}$ attains some extreme value (near 0 or~1), because the shrinkage does not change anything for $\hat{\pi}={\sf E}\pi$.

Table~\ref{tab1} overviews some useful information for three versions of the Bayesian estimator of $\pi$, i.e.~for 3 different choices of the prior distribution of $\pi$, and compares them with 
$\hat{\pi}$. All three of them have the form (\ref{e:gen}) and the table evaluates the corresponding values of $\lambda$ and ${\sf E}\pi$.

\begin{table}
\begin{center}
\caption{Estimators of $\pi$ in the binomial model: the MLE and three Bayesian versions.}
\begin{tabular}{p{13mm}p{28mm}p{25mm}p{22mm}p{15mm}}
\hline
Estimator & Prior & Formula & \multicolumn{2}{c}{Shrinkage (\ref{e:gen})}\\
\hline

\\

MLE & - & $\hat{\pi} = \frac{1}{n} \sum_{i=1}^n X_i$ & $\lambda=1$ & -\\

\\

$\tilde{\pi}_\beta$ & $\mbox{Beta}(a,b)$ & $\tilde{\pi}_\beta = \frac{\sum_{i=1}^n X_i + a}{n+a+b}$ & $\lambda=\frac{n}{n+a+b}$ & ${\sf E}\pi=\frac{a}{a+b}$\\

\\

$\tilde{\pi}_{BL}$ & $U(0,1) = \mbox{Beta}(1,1)$ & $\tilde{\pi}_{BL} = \frac{\sum_{i=1}^n X_i + 1}{n+2}$ & $\lambda=\frac{n}{n+2}$ & ${\sf E}\pi = 1/2$\\

\\

$\tilde{\pi}_J$ & $\mbox{Beta}(1/2,1/2)$ & $\tilde{\pi}_{J} = \frac{\sum_{i=1}^n X_i + 1/2}{n+1}$ & $\lambda=\frac{n}{n+1}$ & ${\sf E}\pi = 1/2$

\\
\hline
\end{tabular}
\label{tab1}
\end{center}
\end{table}

\begin{enumerate}
\item {\bf Beta prior.}
First, let us consider the prior distribution to be $\mbox{Beta}(a,b)$.
Because the two parameters $a>0$ and $b>0$ have to be specified, this prior represents an example of an informative prior.
The estimator corresponds to the MLE obtained with $a+b$ additional experiments, where the success was achieved in $a$~situations.

\item {\bf Uniform prior.}
The prior distribution $\pi \sim U(0,1)$ coincides with $\mbox{Beta}(1,1)$ and is non-informative as it does not depend on any additional parameter. 
As suggested by \cite{tuy}, the prior originally proposed in \cite{bay} is known as the Bayes-Laplace prior \cite{pos}.
The uniform distribution has the largest entropy among all continuous distributions on $[0,1]$ and using it corresponds to the max-entropy principle \cite{den}, i.e. an~important paradigm for the construction of non-informative priors.

\item {\bf Jeffreys prior.}
The prior distribution $\pi \sim \mbox{Beta}(1/2,1/2)$ is the Jeffreys non-informative prior obtained as the square root of the Fisher information of $\pi$, which is equal to $I(\pi) = \left( \pi(1-\pi) \right)^{-1}$.
\end{enumerate}

\begin{table}
\begin{center}
\caption{Important symbols used in Sections~\ref{sec:ber} and \ref{sec:hyp}.}
\begin{tabular}{cc}
\hline
$\pi$ & Probability of the Bernoulli distribution\\
$\hat{\pi}$ & Maximum likelihood estimator of $\pi$\\
$\tilde{\pi}$ & Bayesian estimator of $\pi$\\
$\lambda$ & Regularization parameter\\
$I$ & Fisher information\\
$S$ & Sign test statistic\\
$Z$ & Homogeneity test statistic\\
$T$ & McNemar test statistic\\
\hline
\end{tabular}
\label{tab:notationA}
\end{center}
\end{table}

For the beta prior with parameters $a$ and $b$, choosing values below 1 (i.e.~$a<1$ or $b<1$) can lead to extreme behavior in the posterior, such as a mode at the boundary (0 or 1) when the observed counts are very small. 
This may cause numerical instability or overly concentrated estimates. 
To ensure well-behaved posterior estimates and avoid such issues, we recommend using $a,b \ge 1$, which corresponds to a non-informative or weakly informative prior that provides shrinkage without producing extreme modes.

For all the three choices, the prior distribution is beta and the posterior distribution is also beta \cite{gel}.
The resulting estimator of $\pi$ is obtained as the maximum of the posterior distribution and this may be approximated by~$\hat{\pi}$ for all three choices for $n \to \infty$.
Alternatively to taking the expectation of the posterior distribution, one can also use the mode, resulting in the maximum a~posteriori (MAP) estimator, which tends to be more sensitive to the prior and may provide more stable estimates when data is sparse.

Intensive attention has been paid to the choice of a suitable prior in the Bayesian analysis of categorical data \cite{agr}. 
In \cite{tuy}, the focus was on estimating~$\pi$ for data with extreme outcomes and particularly with zero counts, i.e.~with $X=0$;
naturally, $X=n$ represents an analogous complication. The conclusion was to avoid $a<1$ and $b<1$ for beta priors and thus to avoid Jeffreys prior. Among informative priors for the situation with some prior knowledge about~$\pi$, the beta prior represents a very flexible and highly recommendable choice. The estimates $\tilde{\pi}_{BL}$ and $\tilde{\pi}_J$ are obtained for non-informative prior.
We can say that the concept of ``non-informative'' prior is quite misleading, because the non-informative estimates $\tilde{\pi}_{BL}$ and $\tilde{\pi}_J$ do not correspond to the MLE.

Table~\ref{tab:notationA} summarizes some important notation used throughout Sections~\ref{sec:ber} and \ref{sec:hyp}, including symbols for probabilities, counts, and regularized estimates.
For clarity, we explicitly note that hats denote maximum likelihood estimates, tildes denote Bayesian or shrinkage estimates, and stars denote $\lambda$-regularized statistics. 
This convention is used consistently throughout the paper to keep the notation transparent.

\subsection{Properties of the Bayesian estimators} 
\label{sec:prop}

The following properties of the estimators of Table~\ref{tab1} may be derived in a~straightforward way.
Theorem~\ref{th:2} reveals the shrinkage estimators (\ref{e:gen}) to represent a~compromise between MLE and ${\sf E}\pi$.
Theorem~\ref{th:3} evaluates the maximum possible shrinkage for $\tilde{\pi}_{BL}$ and $\tilde{\pi}_J$.
Theorem~\ref{th:4} compares the shrinkage intensity between~$\tilde{\pi}_{BL}$ and $\tilde{\pi}_J$.

\begin{theorem}
It holds that $\tilde{\pi}_\beta = \hat{\pi} + \mathcal{O} \left(\frac{1}{n}\right)$ for $n\to\infty.$
\label{th:infty}
\end{theorem}

\begin{theorem}
If $X_1,\dots,X_n$ are i.i.d. random variables following the Bernoulli distribution with parameter $\pi$, then it holds for $\gamma \geq 0$ and $\delta>0$ that 
\begin{equation}
\min \left\{ \frac{\sum_{i=1}^n X_i}{n}, \frac{\gamma}{\delta} \right\} \leq \frac{\sum_{i=1}^n X_i+\gamma}{n+\delta} \leq \max \left\{ \frac{\sum_{i=1}^n X_i}{n}, \frac{\gamma}{\delta} \right\}.
\end{equation}
\label{th:2}
\end{theorem}

\begin{theorem} 
If $\hat{\pi}<1/2$, then 
\begin{equation}
\max_{ \hat{\pi} \in [0,1/2)} | \hat{\pi} - \tilde{\pi}_{BL} | = \frac{1}{n+2}
\end{equation}
and this value is attained either for $\hat{\pi}=0$ or for $\hat{\pi}=1$.
If again $\hat{\pi}<1/2$, then 
\begin{equation}
\max_{\hat{\pi} \in [0,1/2)} | \hat{\pi} - \tilde{\pi}_J | = \frac{1}{n+1}
\end{equation}
and this value is attained either for $\hat{\pi}=0$ or for $\hat{\pi}=1$.
\label{th:3}
\end{theorem}

\begin{theorem} 
It holds that
\begin{equation}
\tilde{\pi}_{BL} < \tilde{\pi}_{J} \Longleftrightarrow \hat{\pi}>\frac{1}{2}
\end{equation}
and 
\begin{equation}
\tilde{\pi}_{BL} = \tilde{\pi}_{J} \Longleftrightarrow \hat{\pi}=\frac{1}{2}.
\end{equation}
It already follows that 
\begin{equation}
\tilde{\pi}_{BL} > \tilde{\pi}_{J} \Longleftrightarrow \hat{\pi}<\frac{1}{2}.
\end{equation}
\label{th:4}
\end{theorem}

\section{Hypothesis tests based on Bayesian estimates} 
\label{sec:hyp}

Further, we are interested in the properties of common hypothesis tests for categorical data used with regularized versions of maximum likelihood estimates of the probabilities
Particularly, we consider the sign test for binomial distribution, Pearson $\chi^2$ test of homogeneity, and McNemar test for $2 \times 2$ contingency tables.
The regularized (shrunken) estimates of the probabilities of Section~\ref{sec:ber} will be considered in a~natural way to obtain regularized versions of the test statistics.
The asymptotic distribution of each of the test statistics is derived under the null hypothesis.

\subsection{Sign test}
\label{sec:sig}

We consider the sign test for the binomial distribution assuming $X \sim {\sf Bi}(n,\pi)$. 
We consider the null hypothesis $H_0:~\pi=1/2$. 
The maximum likelihood estimator of~$\pi$ is denoted as $\hat{\pi}=X/n$.
The sign test is based on the test statistic (say $S$) and is performed according to
\begin{equation}
S= 2 \sqrt{n} \left(\hat{\pi} - \frac{1}{2}\right) = \frac{2X-n}{\sqrt{n}} \dist \Omega, \quad \mbox{where}~\Omega \sim {\sf N}(0, 1). 
\label{e:sign}
\end{equation}
This test statistic can be easily derived as the statistic 
$S=(\hat{p}-{\sf E} \hat{p})/\sqrt{{\sf var}\,\hat{p}}$,
where the expectation under $H_0$ is ${\sf E} \hat{p}=1/2$ and the variance under $H_0$ is ${\sf var}\,\hat{p}=1/(4n)$.

\begin{theorem}
Let us assume a random variable $X$ following the binomial distribution ${\sf Bi}(n,\pi)$. If the test statistic $S$ (\ref{e:sign}) exploits the estimate $\pi^*=\lambda X + (1-\lambda) \pi_0$ instead of $\hat{\pi}$, then the resulting test statistic $S$ denoted as $S^*$ fulfills
\begin{equation}
\frac{S^*}{\lambda} \dist \Omega
\end{equation}
under $H_0$, where $\Omega$ follows ${\sf N}(0, 1)$ distribution.
\label{th:sign}
\end{theorem}

The considered penalized estimator $\pi^*=\lambda X + (1-\lambda) \pi_0$ is a Bayesian estimate from Section~\ref{sec:ber} with a particular choice of $\lambda \in [0,1]$.

\subsection{Test of homogeneity} 
\label{sec:indep}

Consider a $2 \times 2$ contingency table, where each of the two columns is assumed to follow a binomial distribution. The Pearson $\chi^2$ test of homogeneity will then be applied to assess whether the distributions across the columns are significantly different.

Let us assume the random samples to be measured in the total number of 2~populations (groups). A~binary variable is observed in each randomly selected unit. 
The observed counts for the two groups are presented in the contingency table
\begin{equation}
\begin{tabular}{l|c|c|c}
& Group A & Group B & $\sum$\\
\hline
Success & $n_{11}$ & $n_{12}$  & $n_{1 \cdot}$\\
Failure & $n_{21}$ & $n_{22}$  & $n_{2 \cdot}$\\
\hline $\sum$ & $n_{\cdot 1}$  & $n_{\cdot K}$ & $n$
\end{tabular}.
\label{tabulka}
\end{equation}
Each column follows a binomial model with parameters $(n_{\cdot j}, p_j)$, where $p_j$ is the probability of success in population $j$. The probabilities are presented in the table
\begin{equation}
\begin{tabular}{l|c|c}
& Group A & Group B \\ \hline
Success & $\pi_1$ & $\pi_2$ \\
Failure & $1-\pi_1$ & $1-\pi_2$ \\ \hline
$\sum$ & 1 & 1 \\
\end{tabular}.
\label{tabulkapsti}
\end{equation}
The maximum likelihood estimates of $\pi_1$ and $\pi_2$ will be denoted by $\hat{\pi}_1$ and $\hat{\pi}_2$, respectively.

We consider the null hypothesis $H_0:~\pi_1=\pi_2$. 
One of alternative ways of describing the $\chi^2$ test for a $2 \times 2$ table is to consider the test statistic 
\begin{equation}
Z= (\hat{\pi}_2 - \hat{\pi}_1) \sqrt{\frac{n n_{\cdot 1} n_{\cdot 2}}{n_{1 \cdot} n_{2 \cdot}}} \dist \Xi,
\label{e:z}
\end{equation}
where $\dist$ denotes convergence of distribution and $\Xi$ follows ${\sf N}(0,1)$ distribution. It holds that $Z^2$ is exactly equal to the $\chi^2$ statistic of Pearson test; unlike $\chi^2$, the test statistic $Z$ may also be used for a one-sided test.


To justify the test statistic $Z$, it is sufficient to show that the square of $Z$ fulfils
\begin{align}
Z^2 &= \left( \frac{n_{11}}{\ntj} - \frac{n_{12}}{\ntd} \right)^2 \frac{n \ntj \ntd}{\njt \ndt}\nonumber\\
&= \left( \frac{\njj \ndd - \njd \ndj}{\ntj \ntd} \right)^2 \frac{n \ntj \ntd}{\njt \ndt}\nonumber\\
&= \frac{n(\njj \ndd - \njd \ndj)^2}{\njt \ndt \ntj \ntd} = \chi^2,
\end{align}
i.e.~$Z^2$ is equal to the Pearson $\chi^2$ test statistic.

\begin{theorem}
We consider a $2 \times 2$ contingency table. If the test statistic $Z$ is considered with estimates
\begin{equation}
\pi_1^*= \lambda \frac{n_{11}}{n_{\cdot 1}} + (1-\lambda) \frac{n_{1 \cdot}}{n} \quad\mbox{and}\quad \pi_2^*= \lambda \frac{n_{12}}{n_{\cdot 2}} + (1-\lambda) \frac{n_{1 \cdot}}{n},
\label{e:regest}
\end{equation}
then the resulting test statistic $Z$ denoted as $Z^*$ fulfills
\begin{equation}
\frac{Z^*}{\lambda} = \frac{(\pi_2^* - \pi_1^*)}{\lambda}\sqrt{\frac{n n_{\cdot 1} n_{\cdot 2}}{n_{1 \cdot} n_{2 \cdot}}} \dist \Xi
\end{equation}
under $H_0$, where $\Xi$ follows ${\sf N}(0,1)$ distribution.
\label{th:indep}
\end{theorem}

The regularized estimates in (\ref{e:regest}) are Bayesian estimates from Section~\ref{sec:ber} with a~particular choice of $\lambda \in [0,1]$. While regularization is typically considered in the context of zero counts in contingency tables, the regularization is justified by the connection to Bayesian thinking here.

We note that the shrinkage target for each group's success probability is taken as the pooled success rate, which is estimated from the data. 
This corresponds to an empirical Bayes approach. Importantly, the asymptotic standard normal limits derived remain valid under this choice.

\subsection{McNemar test}
\label{sec:mcn}

Let us now consider a $2 \times 2$ table following a multinomial model with a fixed total number of samples $n$. McNemar test is a test of symmetry and at the same time a~test of marginal homogeneity \cite{smi}. Often, the test is applied if comparing the effect of a~treatment before some event and after it. The table of observed counts will be denoted by
\begin{equation}
\begin{tabular}{c|cc|c}
& \multicolumn{2}{c|}{After}\\
Before & $A$ & $B$ & $\sum$\\ \hline
$A$ & $n_{11}$ & $n_{12}$ & $n_{1 \cdot}$\\
$B$ & $n_{21}$ & $n_{22}$ & $n_{2 \cdot}$\\  \hline
$\sum$ & $n_{\cdot 1}$ & $n_{\cdot 2}$ & $n$ \\
\end{tabular} .
\end{equation}
The table of the corresponding probabilities has the form
\begin{equation}
\begin{tabular}{c|cc|c}
& \multicolumn{2}{c|}{After}\\
Before & $A$ & $B$ & $\sum$\\ \hline
$A$ & $\pi_{11}$ & $\pi_{12}$ & $\pi_{1 \cdot}$\\
$B$ & $\pi_{21}$ & $\pi_{22}$ & $\pi_{2 \cdot}$\\  \hline
$\sum$ & $\pi_{\cdot 1}$ & $\pi_{\cdot 2}$ & $1$ \\
\end{tabular} .
\end{equation}
The maximum likelihood estimates of the probabilities are denoted as $\hat{\pi}_{ij}= n_{ij}/n$ for $i=1,2$ and $j=1,2$.
The test of $H_0:~\pi_{12}=\pi_{21}$ is based on the statistic
\begin{equation}
T= \frac{n}{\sqrt{n_{12}+n_{21}}} (\pi_{12} - \pi_{21})  \dist \Xi, \quad \mbox{where}~\Xi \sim {\sf N}(0, 1).
\end{equation}
It can be shown easily that the square of $T$ is equal to the commonly used form of the statistic $\chi^2$ of McNemar test, which fulfils
\begin{equation}
\chi^2 = \frac{n_{12}-n_{21}}{n_{12}+n_{21}} \dist \Xi, \quad \mbox{where}~\Xi \sim \chi^2_1.
\end{equation}

Let us now consider regularized estimators
\begin{equation}
\pi_{12}^* = \lambda \hat{\pi}_{12} + (1-\lambda) \tau \quad\mbox{and}\quad \pi_{21}^* = \lambda \hat{\pi}_{21}+ (1-\lambda) \tau
\label{mcn:est}
\end{equation}
for a given $\tau \in [0,1]$ and $\lambda>0$.
The value of $\tau$ may be given by prior knowledge and the estimates in (\ref{mcn:est}) correspond to Bayesian estimates presented in Section~\ref{sec:ber}.

\begin{theorem}
We consider a $2 \times 2$ contingency table. If the test statistic $Z$ is considered with estimates (\ref{e:regest}),
then the resulting test statistic, denoted as $T^*$, satisfies
\begin{equation}
T^*= \frac{n}{\sqrt{n_{12}+n_{21}}} \frac{(\pi_{12}^* - \pi_{21}^*)}{\lambda} \dist \Xi, \quad \mbox{where}~\Xi \sim {\sf N}(0, 1).
\end{equation}
\label{th:mcnemar}
\end{theorem}

The results of this section are intuitive, but analogous results do not seem to be available for larger contingency tables. Additionally, analogous reasoning for continuous data leads to much more complicated outcomes. For example, the two-sample 
$t$-test based on regularized means can be derived to follow a~noncentral $t$-distribution, which presents a more complex situation with less straightforward application due to challenges in estimating the noncentrality parameter.

Given the simplicity of the test statistics, a practical approach for practitioners is to apply a parametric bootstrap using the shrunken probabilities. 
This allows assessment of finite-sample type I error rates and can improve the reliability of hypothesis testing when sample sizes are small.

Practitioners should be aware that the asymptotic standard normal limits for $S^*/\lambda$, $Z^*/\lambda$, and $T^*/\lambda$ under $H_0$ require applying the scaling by $\lambda$ when comparing to standard normal critical values. 
If the unscaled statistics $S^*$, $Z^*$, or $T^*$ are used directly, the critical values must be adjusted accordingly to account for the factor~$\lambda$.

\section{Regularized versions of association measures} 
\label{sec:mea}

Using regularized test statistics, it is possible to obtain alternative versions of various association measures or other characteristics for categorical data.
Let us assume the situation of Section~\ref{sec:hyp} with a $2 \times 2$ contingency table.
To avoid any misunderstanding, the regularized version of the statistic $Z^*$ (\ref{e:zstar}) will now be denoted as~$Z^*(\lambda)$; the value of $\lambda$ is always chosen according to the choices of Table~\ref{tab1}.
Table~\ref{tab:notationB} summarizes some important notation used throughout Sections~\ref{sec:mea} and \ref{sec:ham}.

\begin{table}
\begin{center}
\caption{Important symbols used in Sections~\ref{sec:mea} and \ref{sec:ham}.}
\begin{tabular}{cc}
\hline
$C_P$ & Pearson contingency coefficient\\
$\varphi$ & Phi coefficient\\
$V$ & Cram\'er's coefficient\\
${\sf MI}$ & Mutual information\\
$t_{ij}$ & Shrinkage target for $i=1,\dots,I$ and $j=1,\dots,J$\\
$\tau_i$ & Shrinkage target for $i=1,\dots,I$\\
$\eta_j$ & Shrinkage target for $j=1,\dots,J$\\
\hline
\end{tabular}
\label{tab:notationB}
\end{center}
\end{table}

\begin{theorem}
Let us assume a $2 \times 2$ contingency table with the test statistic~(\ref{e:z}) denoted as $Z_1$.
Let us assume another $2 \times 2$ contingency table with the test statistic~(\ref{e:z}) denoted as $Z_2$.
Assuming a given $\lambda>0$, let regularized versions of $Z_1$ and~$Z_2$ be denoted as $Z_1^*(\lambda)$ and $Z_2^*(\lambda)$, respectively.
If it holds $Z_1>Z_2$, then it holds that $Z_1^*(\lambda) > Z_2^*(\lambda)$.
\label{l:order}
\end{theorem}

Theorem~\ref{l:order} states that a regularized version of (\ref{e:z}) is a suitable basis for regularized association measures. 
The proof follows from Theorem~\ref{th:indep}.

The regularized test statistic $Z^*(\lambda)$ preserves the main invariance 
properties of the original statistic. In particular, it is invariant to 
relabeling of categories and remains monotone in the odds ratio for 
$2\times 2$ contingency tables. 
Moreover, while the regularization introduces a slight bias in small samples, 
this is accompanied by a reduction in variance, in line with the classical 
bias–variance tradeoff.

Let us now consider two commonly used measures of association for contingency tables derived directly from~$Z$~(\ref{e:z}), or in fact from its square, which is the Pearson's $\chi^2$ statistic.
The measures are the Pearson contingency coefficient defined as 
 \begin{equation}
	C_P = \sqrt{\frac{Z^2}{Z^2+n}},
	\end{equation}
and the phi coefficient defined as 	
\begin{equation}
	\varphi = \sqrt{\frac{Z^2}{n}}.
	\label{e:phi}
	\end{equation}	
	
\begin{theorem}
Theorem~\ref{l:order} holds if the regularized statistic $Z^*(\lambda)$ (\ref{e:z}) is replaced by 
\begin{enumerate}
\item the regularized Pearson contingency coefficient defined as
\begin{equation}
C^*_P(\lambda) = \sqrt{\frac{(Z^*(\lambda))^2}{(Z^*(\lambda))^2+n}},
\end{equation} 
\item the regularized phi coefficient defined as
\begin{equation}
	\varphi^*(\lambda) = \sqrt{\frac{(Z^*(\lambda))^2}{n}}.
	\label{e:phireg}
	\end{equation}	
\end{enumerate}
\end{theorem}

For an overview of properties of the association measures, we refer to \cite{agr}.
Another commonly used association measure is Cram\'er's coeffcient $V$. 
We may define its regularized version for an $I \times J$ table in the form
  \begin{equation}
	V= \sqrt{\frac{Z^2}{n(q-1)}}, \quad\mbox{where}\quad q= \min\{I,J\};
	\end{equation}
nevertheless, for a $2 \times 2$ table, $V^*(\lambda)$ coincides with the regularized phi coefficient~(\ref{e:phireg}).


\section{Regularized mutual information} 
\label{sec:ham}

In this section, we introduce a formal definition of a regularized version of mutual information for a two-way contingency table of size $I \times J$. While various regularized versions of mutual information have been employed in the analysis of real-world data 
(e.g.~\cite{kalmrmr}), we aim to provide a more rigorous treatment here. Although the underlying idea is not novel, we believe that offering a more formal and theoretically precise definition is desirable.
Our approach using the Bayesian estimates of Section~\ref{sec:ber} is inspired by \cite{hau}, who formulated a regularized version of the Shannon entropy, where the regularized (shrinkage) version is known as Bayesian entropy.
In Section~\ref{sec:zero}, we derive that the regularized mutual information does not depend on cells with zero counts.

We consider a discrete variable $X$ with categories $\{1,\dots,I\}$ and a discrete variable $Y$ with categories $\{1,\dots,J\}$.
Let us denote the joint probability of the category~$i$ for $X$ and of the category $j$ for $Y$ by $\pi_{ij}$. In this notation, the parameters are $\pi_{11}, \dots, \pi_{IJ}$ and it holds naturally that $\sum_{i=1}^I \sum_{j=1}^J \pi_{ij}=1$.
The mutual information between the two variables $X$ and $Y$ is defined by
\begin{equation}
{\sf MI}(X,Y) = - \sum_{i=1}^I \sum_{j=1}^J \hat{\pi}_{ij} \log \frac{\hat{\pi}_{ij}}{\hat{\pi}_{i \cdot} \hat{\pi}_{\cdot j}}.
\label{e:mi}
\end{equation}

Let $\hat{\pi}_{11},\dots,\hat{\pi}_{IJ}$ denote maximum likelihood estimates of $\pi_{11},\dots,\pi_{IJ}$. 
Let us consider regularized estimates of $\pi_{ij}$ obtained as 
\begin{equation}
\pi^*_{ij} = \lambda \hat{\pi}_{ij} + (1-\lambda) t_{ij}
\label{e:pihat}
\end{equation}
with a given $t_{ij} \in (0,1)$ for all possible $i$ and $j$, where $\lambda \in [0,1]$ is a regularization parameter. 
We assume the shrinkage target $(t_{11},\dots,t_{IJ})^T \in \real^{IJ}$ to be a vector of non-negative values.
We assume also the vectors $\zeta=(\zeta_1,\dots,\zeta_I)^T \in\real^I$ and $\eta=(\eta_1,\dots,\eta_J)^T\in\real^J$ to be vectors of non-negative values.
We assume
\begin{equation}
\sum_{i=1}^I t_{ij}=\eta_j \quad \mbox{for}~j=1,\dots,J, \quad
\sum_{j=1}^J t_{ij}=\zeta_i \quad \mbox{for}~i=1,\dots,I,  
\end{equation}
and $\sum_{i=1}^I \sum_{j=1}^J t_{ij} = 1$.

Using the regularized estimates (\ref{e:pihat}) within the population mutual information~(\ref{e:mi}), we obtain the regularized empirical version of ${\sf MI}(X,Y)$ formally defined by
\begin{equation}
\begin{split}
&\widehat{{\sf MI}}(X,Y) =  -\sum_{i=1}^I \sum_{j=1}^J \left( \lambda \hat{\pi}_{ij} + (1-\lambda) t_{ij} \right)\cdot\\
&\cdot \left[ \log \left( \lambda\hat{\pi}_{ij} + (1-\lambda)t_{ij} \right) - \log \left(\lambda \hat{\pi}_{i \cdot} + (1-\lambda)\zeta_i \right) - \log \left(\lambda\hat{\pi}_{\cdot j} + (1-\lambda)\eta_j \right) \right].
\end{split}
\label{e:mistar}
\end{equation}

We note that the regularized mutual information remains nonnegative and equals zero if and only if $X$ and $Y$ are independent, provided that the vectors $t$, $\zeta$, and $\eta$ used for shrinkage are strictly positive and appropriately normalized. 
This ensures that the regularized MI preserves the fundamental properties of the classical mutual information, giving practitioners confidence in its use.

\subsection{Zero counts}
\label{sec:zero}

We investigate the contribution of zero counts to the value of the regularized mutual information. Zero counts are a common phenomenon in contingency tables with a~large number of rows and/or columns. We show that the mutual information does not actually depend on these zero counts, which may therefore be disregarded in the formula~(\ref{e:mistar}).

\begin{theorem}
We consider a discrete variable $X$ with categories $\{1,\dots,I\}$ and a~discrete variable $Y$ with categories $\{1,\dots,J\}$ with the notation of Section~\ref{sec:ham}.
Let $\hat{\pi}_{i \cdot}>0$ for every $i=1,\dots,I$ and let $\hat{\pi}_{\cdot j}>0$ for every $j=1,\dots,J$.
Let us have vectors of non-negative values $t \in \real^{IJ}$, $\zeta\in\real^I$, and $\eta\in\real^J$.
\begin{enumerate}
\item Let us consider a given pair $[i,j]$, for which $\hat{\pi}_{ij}=0$. Let us assume $t_{ij}=0$.
  Then it holds that 
  \begin{equation}\begin{split}
  &\left( \lambda \hat{\pi}_{ij} + (1-\lambda) t_{ij} \right)\left[ \log \left( \lambda\hat{\pi}_{ij} + (1-\lambda)t_{ij} \right) \right.\\
  &\left. - \log \left(\lambda \hat{\pi}_{i \cdot} + (1-\lambda)\zeta_i \right) - \log \left(\lambda\hat{\pi}_{\cdot j} + (1-\lambda)\eta_j \right) \right] = 0.
  \end{split}
  \end{equation}
\item Let the set of all pairs $[i,j]$, for which $\hat{\pi}_{ij}=0$, be denoted by ${\cal S}.$ Then the regularized empirical mutual information fulfils
  \begin{equation}
  \begin{split}
  \widehat{{\sf MI}}(X,Y) &=  -\sum_{\substack{i=1,\,j=1\\ [i,j]\notin {\cal S}}}^{I,J} \left( \lambda \hat{\pi}_{ij} + (1-\lambda) t_{ij} \right) \left[ \log \left( \lambda\hat{\pi}_{ij} + (1-\lambda)t_{ij} \right)\right.\\
  &\left.  - \log \left(\lambda \hat{\pi}_{i \cdot} + (1-\lambda)\zeta_i \right) - \log \left(\lambda\hat{\pi}_{\cdot j} + (1-\lambda)\eta_j \right) \right].
  \end{split}
  \end{equation}
\end{enumerate}
\label{th:zero}
\end{theorem}

We note that Theorem~\ref{th:zero} assumes strictly positive row and column marginals. 
In practice, if an entire row or column is zero (which can occur in sparse or high-dimensional contingency tables), 
the corresponding categories can be removed from the computation of the regularized mutual information, 
or a small positive value can be imputed to ensure numerical stability. 
Such adjustments maintain the validity of the regularized MI while avoiding undefined logarithms.

\section{Conclusion}  
\label{sec:con}

This paper contributes to the growing body of work at the intersection of Bayesian estimation and categorical data analysis, offering valuable insights for addressing challenges commonly encountered in machine learning applications \cite{kalsjm}. By introducing regularized estimators and a novel uncertainty coefficient for measuring associations between categorical variables, we provide tools that can improve the robustness and interpretability of machine learning models, especially when dealing with high-dimensional data \cite{kalmat}. These methods offer promising directions for future research, particularly in areas such as feature selection, model calibration, and handling small or imbalanced datasets \cite{zha}. We hope this work serves as a source of inspiration for researchers looking to advance statistical methodologies within the machine learning community, opening new avenues for improving model performance and reliability in real-world applications.

Important regularized approaches for the analysis of categorical data can be naturally obtained within the Bayesian framework, which allows combining observed data with prior information. Even in the absence of prior knowledge, it may be advantageous to consider Bayesian (shrinkage) estimates instead of relying solely on maximum likelihood estimates. A further benefit is that Bayesian methods provide the entire posterior distribution of parameters rather than just point estimates, enabling richer uncertainty quantification.

The effect of regularization on hypothesis tests and association measures can be complex to evaluate, with the exception of a few simple models discussed in this paper. In particular, when using regularized probabilities, the asymptotic null distribution of test statistics is modified, and the impact can be assessed in specific cases. Practical guidance on the choice of the regularization parameter $\lambda$ is limited; however, if the variance of the prior distribution were known or could be reliably estimated, $\lambda$ could be chosen to reflect the relative weight of prior information versus data, providing a principled approach for practical applications. Developing systematic rules for selecting $\lambda$ in this way remains an interesting direction for future research.

Future work will focus on developing novel hypothesis tests for approximated neural networks for categorical data with large numbers of parameters, particularly emphasizing high-dimensional contingency tables. These tests will serve as diagnostic tools, such as zero-weight tests, prioritizing interpretability and computational efficiency. Moreover, the insights presented in this paper offer opportunities to enhance machine learning models, enabling them to better handle large, sparse, and complex datasets.

\section*{Appendix: Proofs}


\begin{proof}[of Theorem~\ref{th:infty}]
The result follows from a simple reformulation 
\begin{equation}
\tilde{\pi}_\beta = \hat{\pi} \frac{n}{n+a+b} + \frac{a}{n+a+b} = \hat{\pi} \left(1-\frac{a+b}{n+a+b} \right) + \frac{a}{n+a+b} 
\end{equation}
and subsequently from 
\begin{equation}
\tilde{\pi}_\beta = \hat{\pi} \left(1- \mathcal{O} \left( \frac{1}{n} \right)\right) + \mathcal{O} \left( \frac{1}{n} \right), \quad n \to\infty.
\end{equation}
\end{proof}

\begin{proof}[of Theorem~\ref{th:2}]
The result follow immediately from the following property. Let $\alpha \geq 0$, $\beta>0$, $\gamma \geq 0$, and $\delta>0.$
Then it holds that
\begin{equation}
\min \left\{ \frac{\alpha}{\beta}, \frac{\gamma}{\delta} \right\} \leq \frac{\alpha+\gamma}{\beta+\delta} \leq \max \left\{ \frac{\alpha}{\beta}, \frac{\gamma}{\delta} \right\}.
\end{equation}
\end{proof}

\begin{proof}[of Theorem~\ref{th:sign}]
We may express
\begin{equation}\begin{split}
\frac{1}{\lambda} S^* &= \frac{1}{\lambda} 2 \sqrt{n} \left(\pi^* - \frac{1}{2}\right)\\
&= \frac{1}{\lambda} 2 \sqrt{n} \left( \lambda \frac{X}{n} + (1-\lambda) \frac{1}{2} - \frac{1}{2} \right)\\
&= \frac{1}{\lambda} 2 \sqrt{n} \left( \lambda \frac{X}{n} - \lambda \frac{1}{2} \right)\\
&= S, 
\end{split}\end{equation}
from which the asymptotic distribution of (\ref{e:sign}) immediately follows.
\end{proof}

\begin{proof}[of Theorem~\ref{th:indep}]
The result follows from
\begin{equation}
Z^* = \left(\lambda \frac{n_{12}}{n_{\cdot 2}} - \lambda \frac{n_{11}}{n_{\cdot 1}} \right) \sqrt{\frac{n n_{\cdot 1} n_{\cdot 2}}{n_{1 \cdot} n_{2 \cdot}}} = \lambda Z.
\label{e:zstar}
\end{equation}
\end{proof}

\begin{proof}[of Theorem~\ref{th:mcnemar}]
The result follows from
\begin{equation}\begin{split}
T^* &= \frac{n}{\sqrt{n_{12}+n_{21}}}  \left( \lambda \hat{\pi}_{12} + (1-\lambda)\tau - \lambda \hat{\pi}_{21} -(1-\lambda)\tau \right) \\
&= \frac{n}{\sqrt{n_{12}+n_{21}}}  \left( \lambda \hat{\pi}_{12}  - \lambda \hat{\pi}_{21}  \right) = \lambda T.
\end{split}
\end{equation}
\end{proof}

\section*{Acknowledgements} 

\noindent
The author would like to thank an anonymous referee for valuable suggestions and Pavel Rak and Ane\v zka Falt\'ynkov\'a (both MFF UK) for technical help.

\bibliography{bibKalina}

\end{document}